\documentclass[pre,twocolumn,showpacs, superscriptaddress, eqsecnum,
amsmath,amssymb,superscriptaddress
]{revtex4}
\newcommand{\be}{\begin{equation}}
\newcommand{\ee}{\end{equation}} 
\newcommand{\bea}{\begin{eqnarray}} 
\usepackage{graphicx}
\newcommand{\eea}{\end{eqnarray}}
\usepackage{epsfig}
\usepackage{bm}
\usepackage{amsmath, amsthm, amssymb}
\newtheorem*{thm1}{Theorem 1}
\newtheorem*{thm2}{Theorem 2}

\begin{document}
\title{Log-Poisson Cascade Description of Turbulent Velocity Gradient Statistics}
\author{M. Kholmyansky}
\affiliation{Faculty of Engineering, Tel Aviv University, Tel Aviv 69978, Israel}
\author{L. Moriconi}
\author{R.M. Pereira}
\affiliation{Instituto de F\'\i sica, Universidade Federal do Rio de Janeiro, \\
C.P. 68528, 21945-970, Rio de Janeiro, RJ, Brazil}
\author{A. Tsinober}
\affiliation{Faculty of Engineering, Tel Aviv University, Tel Aviv 69978, Israel}
\affiliation{
Institute for Mathematical Sciences and Department of Aeronautics,
Imperial College, London, 53 Princes Gate, Exhibition Road, South Kensington Campus,
London SW7 2PG, United Kingdom}

\begin{abstract} 
The Log-Poisson phenomenological description of the turbulent energy cascade is 
evoked to discuss high-order statistics of velocity derivatives and the mapping 
between their probability distribution functions at different Reynolds numbers. 
The striking confirmation of theoretical predictions suggests that numerical 
solutions of the flow, obtained at low/moderate Reynolds numbers can play an 
important quantitative role in the analysis of experimental high Reynolds 
number phenomena, where small scales fluctuations are in general inaccessible 
from direct numerical simulations.
\end{abstract}
\maketitle

\section{Introduction}

Since the pioneering experimental work of Batchelor and Townsend, published exactly sixty years 
ago \cite{batch-town} it is known that scale dependent galilean invariant observables, like velocity 
differences, fluctuate in a strongly non-gaussian way at small scales
in turbulent flows. This kind of statistical behavior, generally referred to as ``intermittency", 
indicates that the K41 picture of turbulence \cite{k41a,k41b}, which actually would correspond to
the existence of a homogeneously distributed energy dissipation field \cite{frisch2}, should break down, 
a fact notoriously anticipated by Landau as early as in 1942 \cite{frisch1}. Not less remarkably, 
long before additional breakthrough experiments were performed \cite{ansel_etal}, phenomenological 
models of the energy cascade advanced the conjecture that intermittency should be related to the 
stochastic multiplicative nature of the energy cascade process \cite{k62,ob62}, implying that small 
scale strong fluctuations are, in some sense, fed by the weaker large scale ones. 

The intermittency phenomenon is commonly associated with the anomalous scaling of velocity 
structure functions. A comprehensive description dealing with both anomalous scaling and the 
non-gaussian behavior of intermittent observables is a major challenge of three-dimensional 
turbulence theory \cite{tsinober}. Small scale strong fluctuations are believed to reflect the dynamics of 
coherent structures like vortex filaments. Even though this is a very open problem, a similar 
physical picture is in fact well-established in simpler contexts, as in Burgers turbulence 
\cite{bec}, with shocks playing the role of ``vortices". 

The log-Poisson model \cite{dubrulle,she} yields perhaps the most intriguing description of the turbulent 
multiplicative cascade, since, as it is well-known, it leads to the accurate She-Leveque intermittency 
exponents of velocity structure functions \cite{she-leveque}. The phenomenological work of She and Leveque 
is also of great physical appeal, once it places vortex filaments as a fundamental ingredient in the 
production of intermittency.

We are interested, in this work, to know what the log-Poisson model may tell us about the 
profiles of velocity gradient pdfs. We deal here with two sets of pdfs for flows associated 
to different Reynolds numbers. One of them is obtained from an atmospheric surface layer 
experiment \cite{guli1_etal, guli2_etal, guli3_etal} and the other from a direct numerical simulation 
(DNS) of homogeneous and isotropic turbulence \cite{donzis}. The underlying motivation in this choice 
of systems is to show that numerical low/moderate Reynolds number results can be useful in the modelling 
of flows that cannot be directly simulated (even in a foreseeable future). The very same claim was put 
forward in a previous letter \cite{kholmy}, where, despite the force of evidence, lacked some 
phenomenological basis, which, then, we develop here. We find that a bridge between 
low and high Reynolds number pdfs can be built within the framework of the log-Poisson model
\cite{comment}.

This paper is organized as follows. In Sec. II we briefly review the multiplicative cascade models,
introduce the log-Poisson model and compute hyperflatness factors of velocity gradient fluctuations, 
comparing them to recent estimates. Two relevant theorems related to velocity gradient 
pdfs are also established. In Sec. III, we present the experimental and numerical data that 
was analysed. In Sec. IV, the experimental and the numerical velocity gradients are closely matched with 
the help of a Monte-Carlo procedure based on the theorems of Sec. II. In Sec. V, we summarize our results 
and point out directions of further research.

\section{Velocity Gradient Statistics}

{\it{ Multiplicative Cascade Models}}
\vspace{0.5cm}

In the multiplicative cascade models \cite{frisch2}, one assumes that energy flows from the integral
scale $L$ to the dissipative scale $\eta$ through a number of ``quantum" steps
associated to eddies of sizes $L,L/a, L/a^2, ...$, where
$a>1$ is an arbitrary rescaling factor. At length scale $\ell_m \equiv L/a^m$ the 
fluctuating energy transfer rate is defined as 
\be
\epsilon_m = \epsilon_0 W_1W_2...W_m \ , \ \label{eq2.1}
\ee
where the $W$'s are positive independent random variables, with unit expectation value, 
$\langle W \rangle =1$, so that the mean energy transfer rate is conserved along the cascade
process, i.e., $\langle \epsilon_m \rangle = \epsilon_0$. The scaling behavior of velocity
structure functions, $S_q(r) \equiv \langle (\delta v)^q \rangle$ is, then, derived with the 
help of Kolmogorov's refined similarity hypothesis, which postulates that fluctuations of 
$\delta v$ at scale $\ell_m$ have the same moments (up to constant numerical factors) as 
$(\epsilon_m \ell_m)^{1/3}$.

Analogous phenomenological arguments can be put forward to deal with the case of 
velocity derivatives -- generically denoted in the following by $\partial v$. The
essential idea is to assume that spatial fluctuations of the velocity field are smooth at 
the dissipative scale, and, therefore,
\be
\partial v \sim \frac{\delta v_\eta}{\eta} \sim (\epsilon_\eta)^{1/3} \eta^{-2/3} \ , \ 
\label{rvelgrad1}
\ee
where, above, $\delta v_\eta$ is the velocity increment defined at length scale $\eta$.
One may write, based on purely dimensional grounds, $\eta \sim (\nu^3 / \epsilon_\eta)^{1/4}$. 
Thus, substituting the latter on (\ref{rvelgrad1}), we get
\be
\partial v \sim \sqrt{\epsilon_\eta / \nu} \ , \ 
\label{rvelgrad2}
\ee
a statistical correspondence not unknown to the previous literature \cite{wyn-tenn}.
A more interesting formulation of the refined similarity hypothesis is given in terms of
probability distributions. As it is clear, velocity gradient pdfs can be always written as
\be
\rho(\partial v ) = \int_0^\infty d \epsilon \rho_1( \epsilon) \rho_2(\partial v | \epsilon) \ , \ 
\label{rho-dv}
\ee
where $\rho_2(\partial v | \epsilon)$ is the velocity gradient pdf conditioned on the energy transfer
rate $\epsilon_\eta = \epsilon$ and $\rho_1( \epsilon)$ is the pdf associated to events which have
$\epsilon_\eta = \epsilon$. The refined similarity hypothesis is, then, the statement that at large
Reynolds numbers,
\be
\rho_2(\partial v | \epsilon) =\sqrt{\nu/ \epsilon} F(\sqrt{\nu/ \epsilon} \partial v) \ , \ 
\label{re_sim_hyp}
\ee
where $F(\cdot)$ is a universal (Reynolds number independent) function of its argument. In fact, taking 
(\ref{rho-dv}) and (\ref{re_sim_hyp}), it is not difficult to show, in agreement with (\ref{rvelgrad2}),
that
\be
\langle (\partial v)^q \rangle = C_q \langle (\epsilon_\eta /\nu)^\frac{q}{2} \rangle \ , \ \label{struct-func}
\ee
where
\be
C_q = \int_{- \infty}^\infty dx x^q F(x) \ . \
\ee

It is worth noting that the form of the universal functions $F(x)$ for the case of velocity 
differences has been the subject of experimental research \cite{gagne,naert}.
As a first approximation, $F(x)$ turns to have a gaussian profile, but one expects asymmetric
corrections to be relevant in the problem of longitudinal structure functions, due to
their non-vanishing skewness.

\vspace{0.5cm}
{\it{Log-Poisson Model}}
\vspace{0.5cm}

In the log-Poisson model \cite{dubrulle,she} one writes down the energy transfer 
rate factors as
\be
W = a^{\mu - m} \ , \ \label{eq2.8}
\ee
where $a=3/2$, $\mu=2/3$ and $m \geq 0$ is a Poisson random variable,
with expectation value 
\be
c= \frac{a \mu}{a-1}\ln a = 2 \ln \left ( \frac{3}{2} \right ) \ . \
\ee

In order to cope with velocity gradient fluctuations, it is necessary to set up in first 
place the total number $N$ of cascade steps associated to the turbulent flow under scrutiny. 
In other words, we would like to find $N$, such that $\eta = L/a^N$. We stress that the 
multiplicative cascade description addressed here is far from being a rigorous framework, since 
we take the Kolmogorov scale $\eta \sim \epsilon_\eta^{-1/4}$ to be a fluctuacting quantity. 
Thus, $N$ should be defined, necessarily, from some averaging procedure. We adopt a simple 
prescription based on the definition of the Reynolds number as \cite{frisch2}
\be
R_e = \frac{L^\frac{4}{3} \epsilon_0^\frac{1}{3}}{\nu} = \left [ \langle \left ( \frac{L}{\eta} \right )^4 \rangle \right ]^\frac{1}{3}
\equiv a^{\frac{4}{3}N} \ . \
\ee
Therefore, we find
\be
N = \frac{3}{4} \log_a R_e \ . \
\ee
An alternative and useful expression for $N$ can be given in terms 
of the Taylor-based Reynolds number $R_\lambda$, which follows by taking the
homogeneous isotropic result $R_\lambda = \sqrt{15 R_e}$ \cite{frisch2},
\be
N = \frac{6}{4}\log_a R_\lambda  - \frac{3}{4}\log_a 15 \ . \
\ee
\vspace{0.5cm}

{\it{Hyperflatness Factors}}
\vspace{0.5cm}

As a direct application of the log-Poisson model, we compute the
Reynolds-dependent velocity gradient hyperflatness factors, 
defined as
\be
H_q (R_\lambda) \equiv \frac{\langle (\partial v)^q \rangle}{\langle (\partial v)^2 \rangle^\frac{q}{2}} \ . \ \label{hyperf}
\ee
A straightforward manipulation of (\ref{hyperf}), taking into account (\ref{eq2.1}), (\ref{struct-func}) and (\ref{eq2.8}), gives
\be
H_q (R_\lambda) = \frac{C_q}{C_2^\frac{q}{2}}
\langle \left ( \frac{\epsilon_\eta}{ \epsilon_0} \right ) ^\frac{q}{2} \rangle  = 
\frac{C_q}{C_2^\frac{q}{2}}
\langle W^\frac{q}{2} \rangle^N
=
A_q R_\lambda^{\alpha_q}\ , \ \label{hyperfb}
\ee
where 
\be
A_q = \frac{C_q}{C_2^\frac{q}{2}} 15^{-\frac{\alpha_q}{2}} \ , \
\ee
with
\be
\alpha_q = \frac{3}{2} \log_a \langle W^{\frac{q}{2}} \rangle = \frac{3}{4} q \mu - \frac{3}{2} \frac{a \mu}{a-1}[ 1 - a^{-\frac{q}{2}}] \ . \ 
\label{eq2.16}
\ee
In particular, the skewness and flatness coefficients predicted by (\ref{eq2.16}) are $\alpha_3 \simeq 0.13$ and $\alpha_4 = 1/3$, respectively. 
Good support is found from the recent account of Ishihara et al. \cite{ishi_etal}, which 
yields $\alpha_3 = 0.11 \pm 0.01$ and $\alpha_4 = 0.34 \pm 0.03$.

If $R_A$ and $R_B$ are Taylor-based Reynolds numbers, respectively associated to flows with $N_A$ and $N_B$ cascade steps, 
then (\ref{hyperfb}) implies that
\be
\frac{H_q(R_A)}{H_q(R_B)} = \langle W^\frac{q}{2} \rangle^{N_A-N_B} \ , \
\ee
and, thus, taking into account (\ref{eq2.16}),
\be
N_A-N_B = \frac{3}{2 \alpha_q} \log_a \frac{H_q(R_A)}{H_q(R_B)} \ , \ \label{eq2.18}
\ee
a quantity that measures the ``distance" between cascades, going to play an important role
in Sec. IV.
\vspace{0.5cm}

{\it{Velocity Gradient PDFs}}
\vspace{0.5cm}

We are interested to explore further consequences of the log-Poisson cascade picture in the setting of velocity gradient pdfs. 
In order to render the exposition more systematic, we introduce two important results in the form of theorems.

\begin{thm1}  

Let $\sigma^2 \equiv \langle (\partial v)^2 \rangle$. The standardized pdf
$\tilde \rho (\partial v) \equiv \sigma \rho( \sigma \partial v)$ has a universal profile
at fixed $R_\lambda$.

\end{thm1}
\begin{proof}

We obtain, from (\ref{rho-dv}) and (\ref{re_sim_hyp}),
\bea
&&\tilde  \rho( \partial v) = \sigma \rho( \sigma \partial v) = \nonumber \\
&=& \sigma \int_0^\infty d \epsilon \rho_1(\epsilon) \sqrt{\nu/ \epsilon} F( \sigma \sqrt{\nu/ \epsilon} \partial v) \nonumber \\
&=&  \nu \sigma^2 \int_0^\infty d \epsilon \rho_1(\nu \sigma^2 \epsilon) \sqrt{1/ \epsilon} F(\sqrt{1/ \epsilon} \partial v) \ . \
\eea
Our task, thus, is to show that $\nu \sigma^2 \rho_1(\nu \sigma^2 \epsilon)$ is indeed universal. Since the sum of Poisson random variables is
also a Poisson random variable, Eqs. (\ref{eq2.1}) and (\ref{eq2.8}) lead, for a cascade with $N$ steps, to
\be
\epsilon_\eta = \epsilon_0 a^{N \mu - m} \ , \
\ee
where $m$ is a Poisson random variable with expectation value $Nc$. We may write, thus,
\be
\rho_1 (\epsilon) = \sum_{m=0}^\infty \frac{ (Nc)^m e^{-Nc}}{m!} \delta( \epsilon - \epsilon_0 a^{N \mu - m}) \ . \
\ee
Now, according to (\ref{struct-func}) we write the variance of $\partial v$ as $\sigma^2 = \epsilon_0 C_2 /\nu$, and, therefore, find
\be
\nu \sigma^2 \rho_1 (\nu \sigma^2 \epsilon) = C_2 \sum_{m=0}^\infty \frac{ (Nc)^m e^{-Nc}}{m!} \delta( C_2 \epsilon - a^{N \mu - m}) \ , \ \label{eq2.22}
\ee
which, in fact, ultimately depends only on $R_\lambda$.

\end{proof}
\begin{thm2} 

Let $A$ and $B$ denote flows with Taylor-based Reynolds numbers $R_A$ and $R_B$,
associated to log-Poisson cascades with $N_A$ and $N_B$ steps, and velocity gradient pdfs
\bea
&&\rho_A(\partial v ) = \int_0^\infty d \epsilon \rho^A_1( \epsilon)
\sqrt{\nu/ \epsilon} F(\sqrt{\nu/ \epsilon} \partial v) \ , \ \nonumber \\
&&\rho_B(\partial v ) = \int_0^\infty d \epsilon \rho^B_1( \epsilon)
\sqrt{\nu/ \epsilon} F(\sqrt{\nu/ \epsilon} \partial v) \ . \
\eea
It follows that 
\be
\tilde \rho_A (\partial v) = \int_0^\infty \frac{dx}{x} K(x) \tilde \rho_B (\frac{\partial v}{x}) \ , \ \label{eq2.24}
\ee
where $K(x)$ is the pdf of the random variable
\be
x =  a^{\frac{1}{2} (N_A-N_B) \mu-\frac{m}{2}}  \ , \ \label{eq2.25}
\ee
which, on its turn, is defined in terms of $m$, a random Poisson variable with 
expectation value $(N_A-N_B)c$.
\end{thm2}
\begin{proof}

A proof follows by direct substitution of the explicit form of $K(x)$ in (\ref{eq2.24}).
Defining $g = a^{N\mu /2}$, with $N=N_A-N_B$, we may write
\be
K(x) = \sum_{m=0}^\infty \frac{ (Nc)^m e^{-Nc}}{m!} \delta( x - g a^{- \frac{m}{2}}) \ . \ \label{eq2.26}
\ee
Using (\ref{eq2.22}) and (\ref{eq2.26}), we obtain, for the RHS of (\ref{eq2.24}),
\bea
&&\int_0^\infty \frac{dx}{x} K(x) \tilde \rho_B (\frac{\partial v}{x}) \nonumber \\
&=& \int_0^\infty \frac{dx}{x} \sum_{m=0}^\infty \frac{ (Nc)^m e^{-Nc}}{m!} \delta( x - g a^{- \frac{m}{2}}) \nonumber \\
&\times& \int_0^\infty d \epsilon \nu \sigma_B^2 \rho_1^B( \nu \sigma_B^2 \epsilon) \sqrt{1 /\epsilon} F(\sqrt{1 / \epsilon}
\partial v / x) \nonumber \\
&=&  C_2 \int_0^\infty \frac{dx}{x} \int_0^\infty d \epsilon \sum_{m=0}^\infty \sum_{m'=0}^\infty \frac{ (Nc)^m e^{-Nc}}{m!} \nonumber \\
&\times& \frac{ (N_Bc)^{m'} e^{-N_Bc}}{m'!} \delta( x - g a^{- \frac{m}{2}}) \delta( C_2 \epsilon - a^{N_B \mu - m}) \nonumber \\
&\times& \sqrt{1 /\epsilon} F(\sqrt{1 / \epsilon} \partial v / x) \ . \  \label{eq2.27}
\eea
Performing the substitution $\epsilon \rightarrow \epsilon/x^2$ in (\ref{eq2.27}) and subsequently integrating over $x$, we get
\bea
&&C_2 \int_0^\infty d \epsilon \sum_{m=0}^\infty \sum_{m'=0}^\infty \frac{ (Nc)^m (N_Bc)^{m'} e^{-N_Ac}}{m!m'!} \nonumber \\
&\times& \delta (C_2 \epsilon - g^2 a^{N_B \mu - m - m'}) \sqrt{1 / \epsilon} F(\sqrt{1/ \epsilon} \partial v) \ . \ \nonumber \\
\label{eq2.28}
\eea
Define, now, $p=m+m'$, so that (\ref{eq2.28}) becomes
\bea
&&C_2 \int_0^\infty d \epsilon \sum_{p=0}^\infty \sum_{m=0}^p \frac{ (Nc)^m (N_Bc)^{m-p} e^{-N_Ac}}{m!(m-p)!} \nonumber \\
&\times& \delta (C_2 \epsilon - g^2 a^{N_B \mu - p}) \sqrt{1 / \epsilon} F(\sqrt{1 / \epsilon} \partial v) \nonumber \\
&=& 
C_2 \int_0^\infty d \epsilon \sum_{p=0}^\infty \frac{ (N_Ac)^p}{p!} e^{-N_Ac}
\delta (C_2 \epsilon - a^{N_A \mu - p}) \nonumber \\
&\times& \sqrt{1 / \epsilon} F(\sqrt{1/ \epsilon} \partial v) = \tilde \rho_A (\partial v) \ . \ 
\eea
\end{proof}

In view of Theorem 2 we can devise a straightforward Monte-Carlo integration procedure in order
to relate velocity-gradient pdfs defined at different Reynolds numbers. In fact, if $x > 0$ and $y$ are
random variables of two independent stochastic process, described, respectively, by pdfs $K(x)$ and 
$\tilde \rho (y)$, then the random variable $z=xy$ is given by the pdf
\bea
\langle \delta(z-xy) \rangle_{x,y} &=& \int dx dy K(x) \tilde \rho(y) \delta(z-xy) \nonumber \\
&=& \int_0^\infty \frac{dx}{x} K(x) \tilde \rho( \frac{z}{x}) = \tilde \rho(z)
\ , \
\eea
where we have used (\ref{eq2.24}) in the last equality above. We have found that it is greatly 
advantageous to use Monte-Carlo integration, instead of more traditional numerical methods, a 
fact probably due to the bad convergence properties of the latter in our particular problem.

\section{atmospheric surface Layer Experiment}

Atmospheric surface layer velocity fluctuations were studied over a grass-covered flat surface in the Sils-Maria valley, Switzerland \cite{guli1_etal}, a place which hosts reasonably stable winds. The results reported in this work correspond to measurements of all of the nine components of the velocity gradient tensor, performed in a tower 3.0 m high. The velocity signal was recorded at sampling rate of $10$ KHz (which was high enough to resolve the dissipative scales), with the help of a 20 hot-wire probe anemometer, specifically designed for the particularities of the field experiment.

\begin{figure}[tbph]
\includegraphics[width=9cm, height=7.5cm]{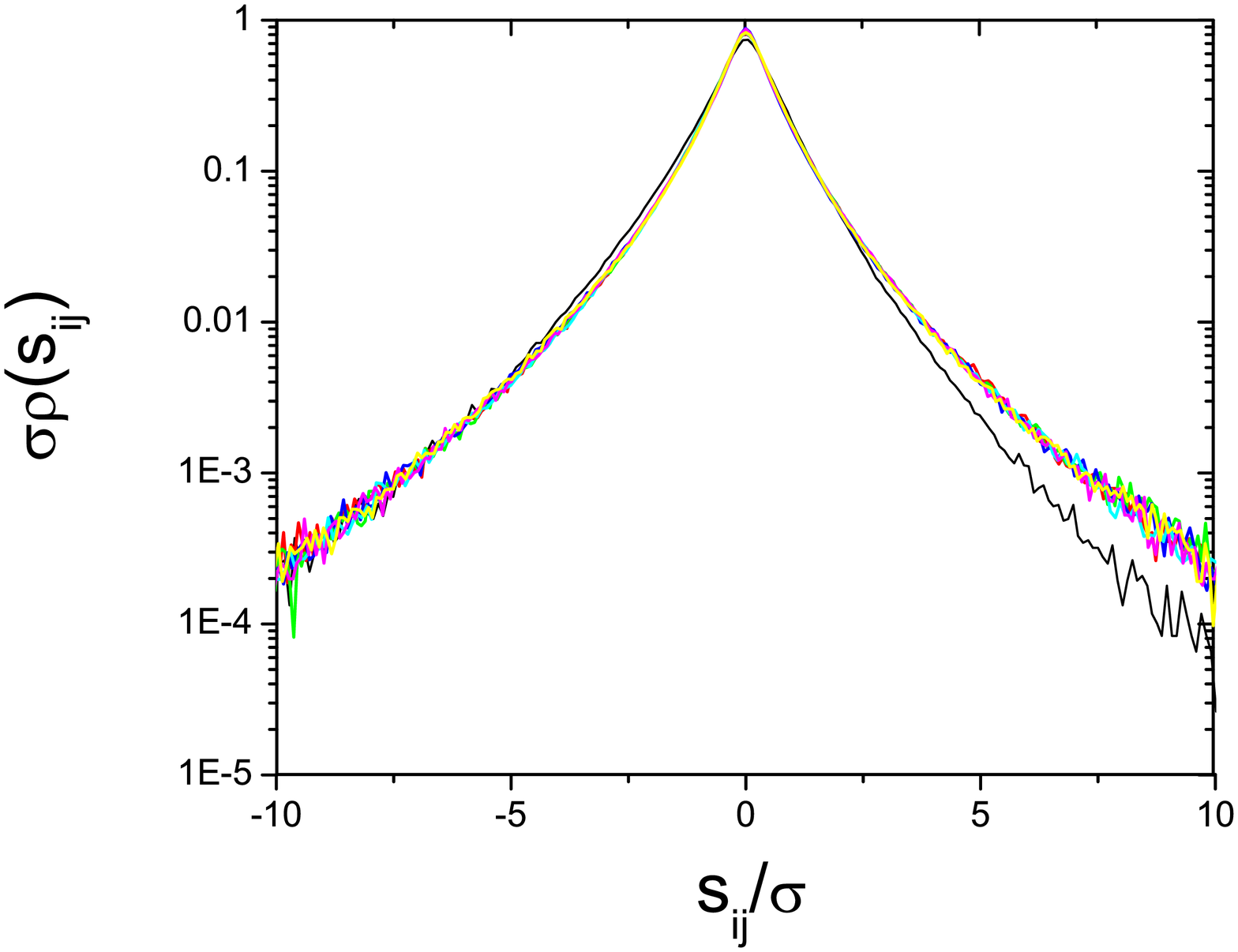}
\caption{Experimental velocity gradient pdfs for
atmospheric surface layer flow with $R_\lambda = 3.4 \times 10^3$. 
Black line: $s_{11}$; collored lines: $s_{ij}$, with $i \neq j$.}
\label{fig1}
\end{figure}
\begin{figure}[tbph]
\includegraphics[width=9cm, height=7.5cm]{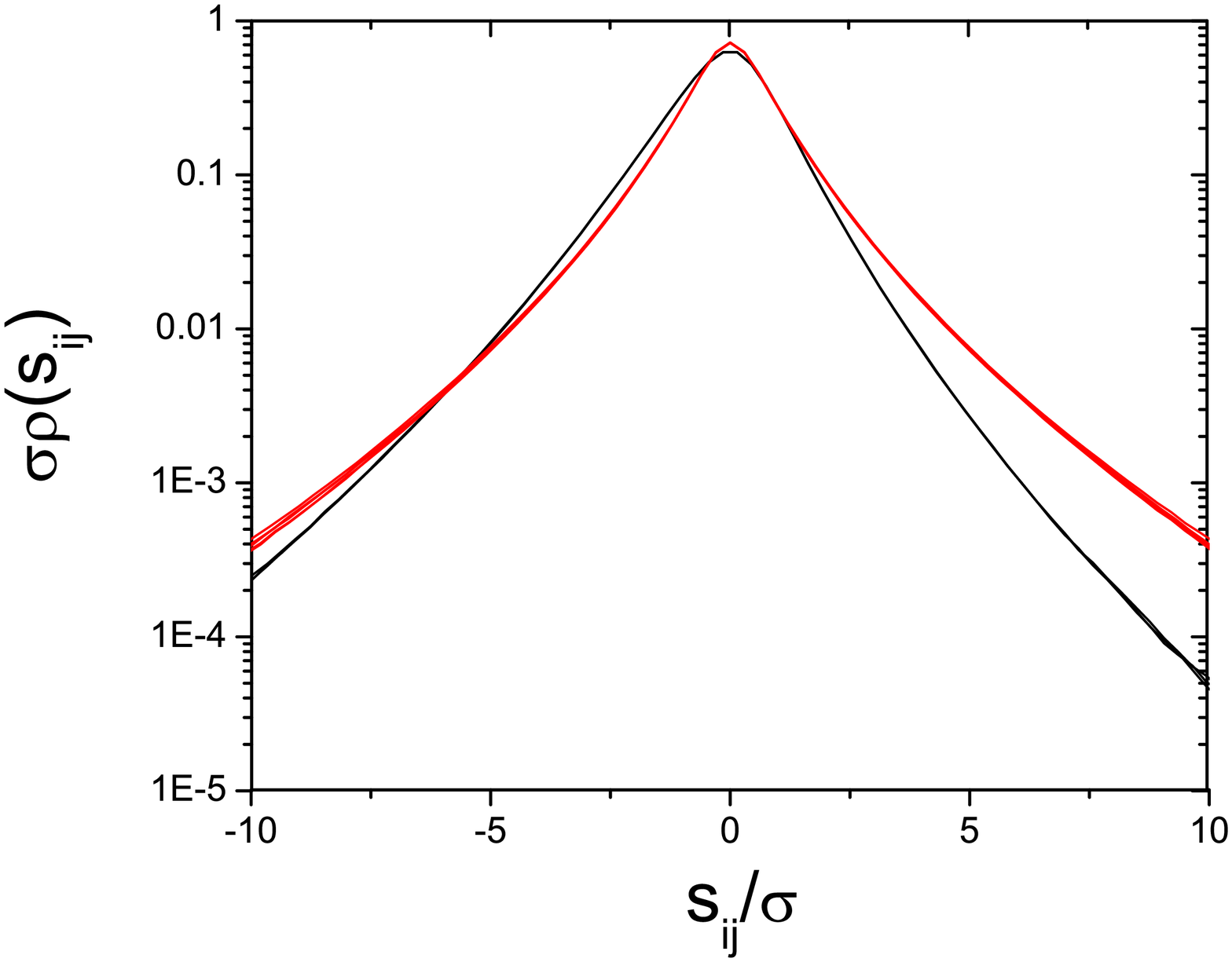}
\caption{Numerical velocity gradient pdfs for
homogeneous isotropic turbulence with $R_\lambda = 240$.
Black lines: diagonal components $s_{ii}$;
red lines: non-diagonal components $s_{ij}$.
}
\label{fig2}
\end{figure}
\begin{figure}[tbph]
\includegraphics[width=9cm, height=7.5cm]{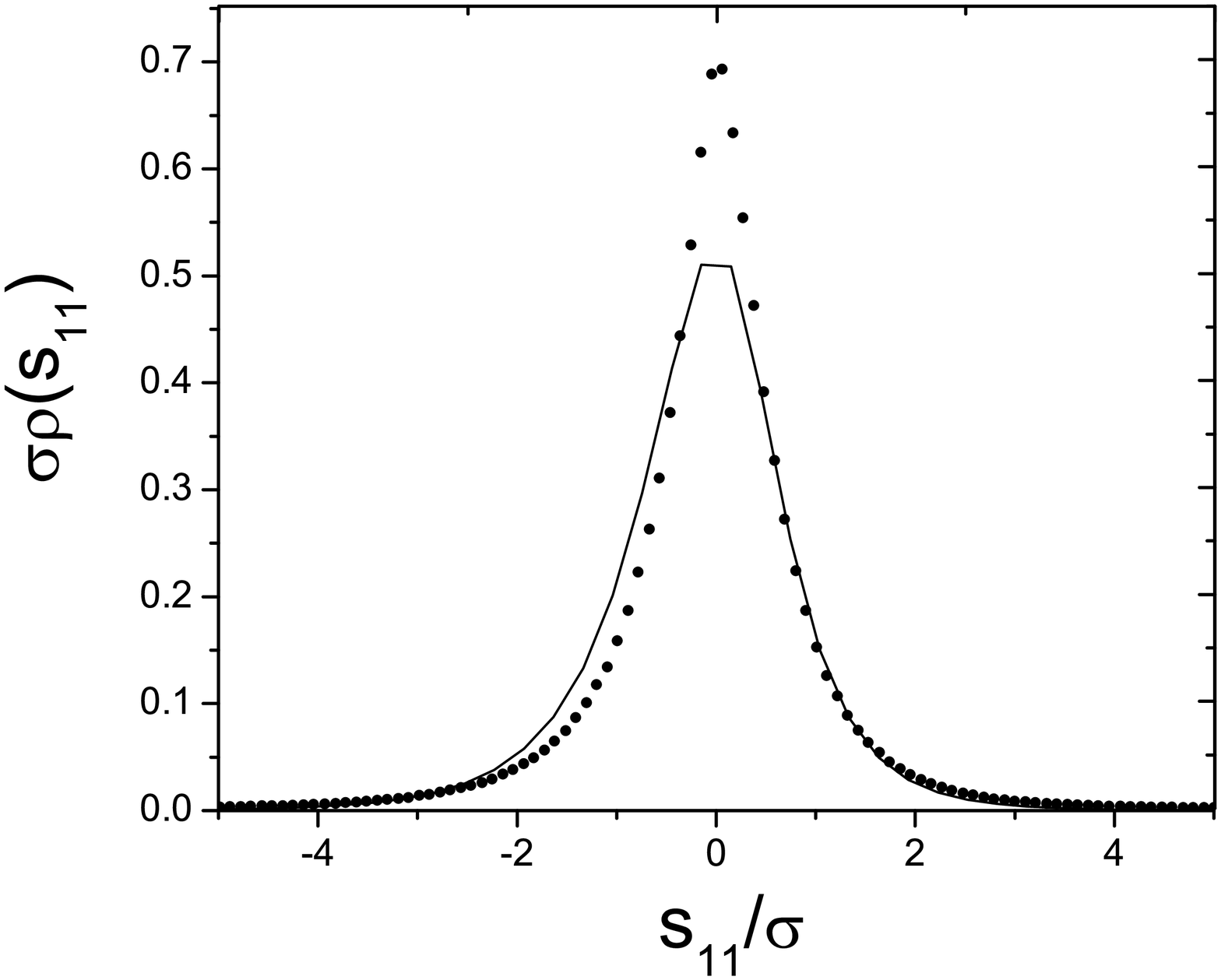}
\caption{Comparison between the numerical ($R_\lambda = 240$;
solid line) and the experimental ($R_\lambda = 3.4 \times 10^3$; dots)
pdfs of $s_{11}$.}
\label{fig3}
\end{figure}
\begin{figure}[tbph]
\includegraphics[width=9cm, height=7.5cm]{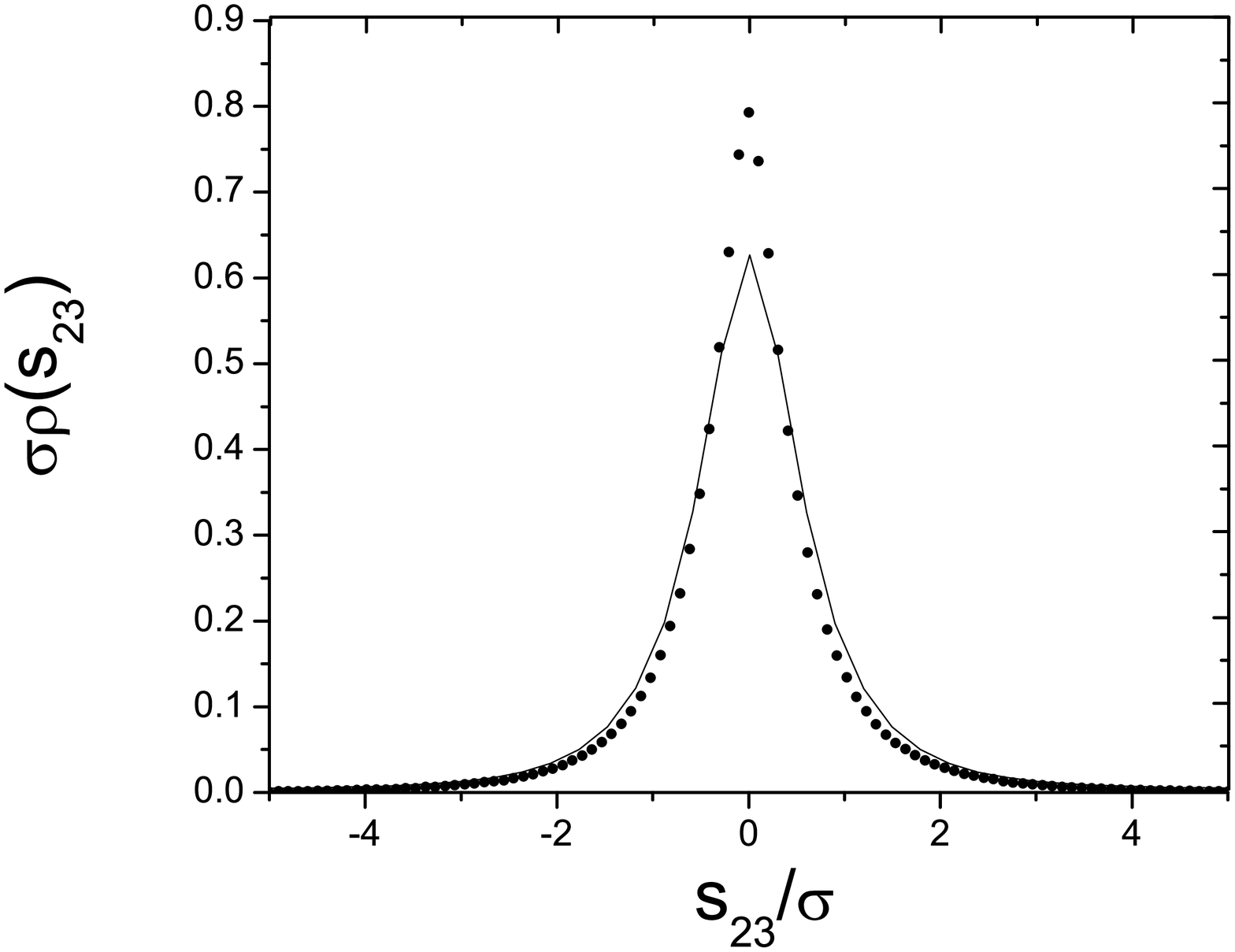}
\caption{Comparison between the numerical ($R_\lambda = 240$;
solid line) and the experimental ($R_\lambda = 3.4 \times 10^3$; dots)
pdfs of $s_{23}$.}
\label{fig4}
\end{figure}

Velocity gradients were computed without resort to the Taylor's frozen turbulence hypothesis. The Taylor-based Reynolds number of the flow, estimated from the Taylor length $\lambda = \sqrt{u_1^2/\langle (\partial_1u_1)^2 \rangle}$ is $R_\lambda = 3.4 \times 10^3$ ($u_1$ is the projection of the velocity fluctuations along the flow direction). We note that since the flow is somewhat anisotropic, the definition of a meaningful Taylor-based Reynolds number may be problematic. We will get back to this point in Sec. IV.

The experimental velocity gradient pdfs are shown Fig.1. We find a good (within error bars) collapse of standardized pdfs of velocity gradients $s_{ij} = \partial_j u_i$ with $i \neq j$. Due to anisotropy effects in the surface layer, however, there is no collapse for the standardized pdfs of diagonal components, $s_{ii}$, and we have discarded the curves for $s_{22}$ and $s_{33}$ assuming, as a working hypothesis to be tested {\it{a posteriori}}, that isotropic results would correspond to the set $\{ s_{11}, s_{ij} \}$, with $i \neq j$.

The central aim of this work is to model the pdfs depicted in Fig. 1 using direct numerical simulation (DNS) results for homogeneous and isotropic turbulence obtained at the considerably lower Taylor-based Reynolds number $R_\lambda = 240$ (the numerical data corresponds to simulations discussed in Ref. \cite{donzis}). The corresponding DNS velocity gradient pdfs are shown in Fig. 2. As it follows from this figure, the pdfs collapse into two distinct groups, associated to the diagonal and non-diagonal components of the velocity gradient tensor $s_{ij}$. Of course, we do not expect that the pdfs given in Fig. 2 yield a direct fitting to the ones of Fig. 1 -- there is a clear discrepancy as shown in Figs. 3 and 4.

\section{Monte-Carlo PDF Reconstruction}

Our computational strategy is to consider the experimental ($R_\lambda = 3.4 \times 10^3$) and the numerical ($R_\lambda = 240$) flows
discussed in Sec. III as the systems $A$ and $B$, respectively, of Theorem 2.  An important parameter here is the cascade distance $N_A-N_B$ 
of these flows. This quantity can be computed by measuring the flatness factors $H_4$ of flows $A$ and $B$ and using them as input parameters in 
(\ref{eq2.18}). 
From the pdfs of $s_{11}$, we get
\bea
H_4(A) &=& 11.5 \ , \ \nonumber \\
H_4(B) &=& 6.6 \ . \
\eea
Therefore, using (\ref{eq2.18}), with $\alpha_4 = 1/3$, we find
\be
N_A - N_B = \frac{9}{2} \log_\frac{3}{2} \left ( \frac{11.5}{6.6} \right ) = 6.16 \ . \
\ee
Due to the discrete structure of the cascade in the multiplicative models, we
take $N_A-N_B = 6$ in the following considerations. 

\begin{figure}[tbph]
\includegraphics[width=9cm, height=7.5cm]{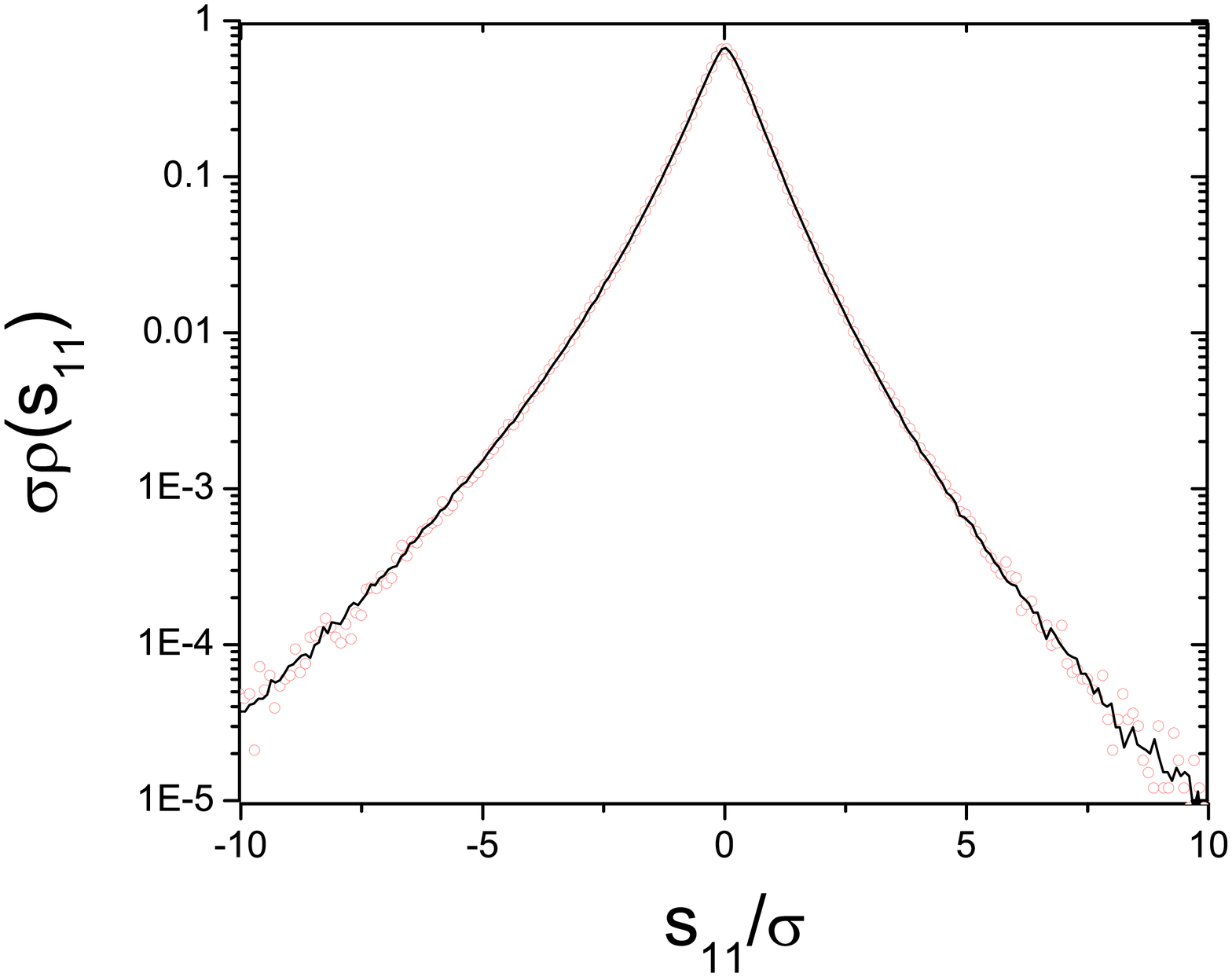}
\caption{The numerically reconstructed pdf of $s_{11}$
(black solid line) is compared to the experimental pdf
(red circles).}
\label{fig5}
\end{figure}
\begin{figure}[tbph]
\begin{center}
\includegraphics[width=9cm, height=7.5cm]{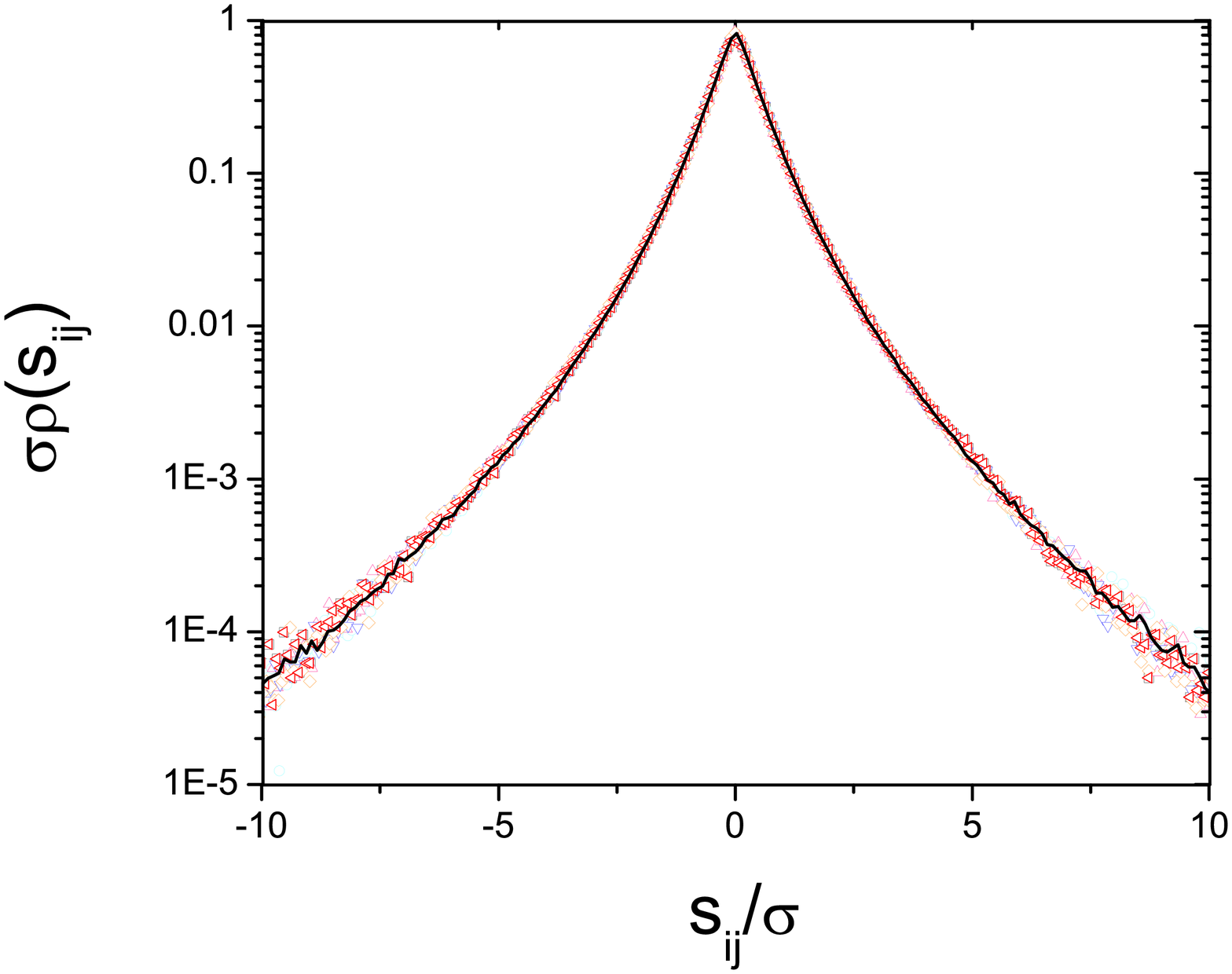}
\end{center}
\caption{The numerically reconstructed pdf of $s_{23}$ 
(black solid line) is compared to the experimental pdfs 
of $s_{ij}$, with $i \neq j$ (collored symbols).}
\label{fig6}
\end{figure}

Using a random Poisson variable generator as the one given in Ref. \cite{knuth}, it is 
straightforward to establish a stochastic process with random variable given by (\ref{eq2.25}). On the 
other hand, in order to generate a stochastic process with random variable described by the 
numerical pdf of $s_{11}$, we proceed in two steps: first, we define an accurate polynomial 
fitting to the $\log_{10} \tilde \rho_B(s_{11})$ profile; second, the polynomial analytical 
distribution just obtained is used in a Monte-Carlo accept-reject algorithm \cite{mc}, 
which produces random variables distributed according to $\tilde \rho_B(s_{11})$.  Analogous 
computations are performed for the numerical pdfs of $s_{23}$, which are taken as a 
representative of the non-diagonal components of the velocity gradient tensor. By multiplying 
the stochastic processes associated to the Poisson and the numerical pdfs we get standardized 
pdfs which would hopefully fit the experimental curves. We have taken a process with 
$2 \times 10^7$ elements. In fact, an excellent agreement is attained from the Monte-Carlo 
reconstructed pdfs, as shown in Figs. 5 and 6. A comparison between the modelled and the 
experimental pdfs is also shown in Figs. 7 and 8 in linear scales, to be contrasted to 
Figs. 3 and 4.

\begin{figure}[tbph]
\includegraphics[width=9cm, height=7.5cm]{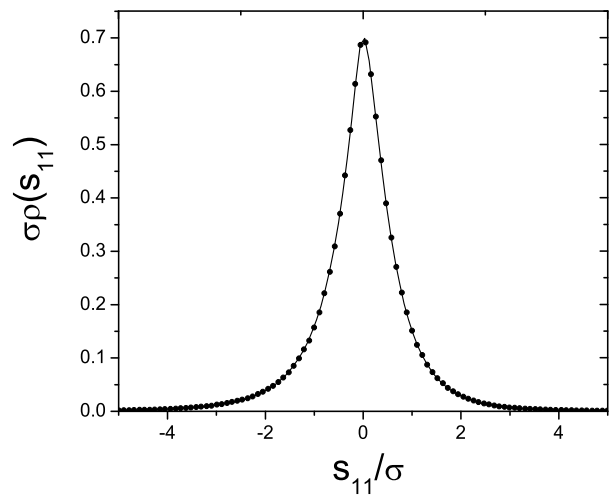}
\caption{The numerically reconstructed pdf of $s_{11}$
(solid line) is compared to the experimental pdf (dots)
in linear scales.}
\label{fig7}
\end{figure}
\begin{figure}[tbph]
\includegraphics[width=9cm, height=7.5cm]{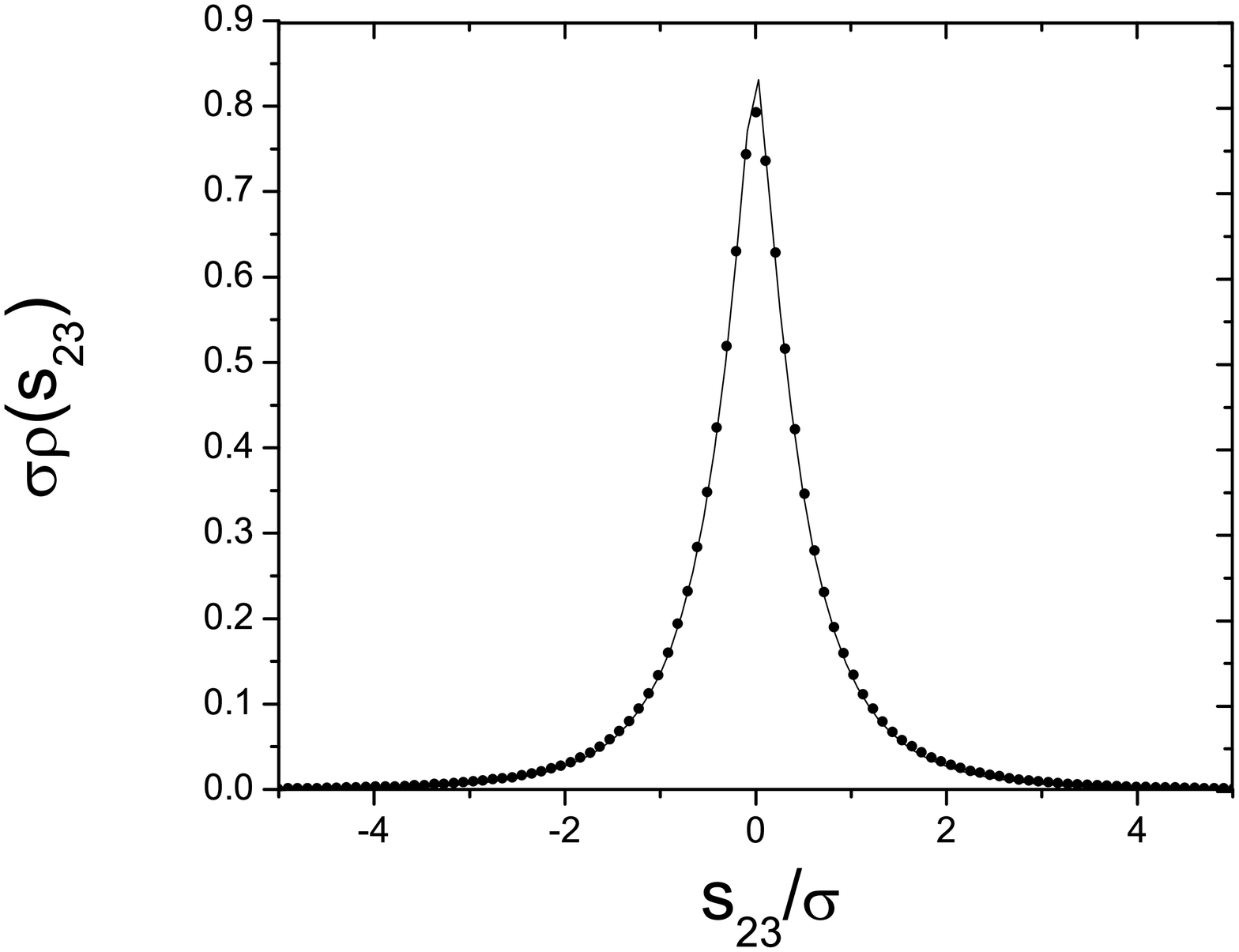}
\caption{The numerically reconstructed pdf of $s_{23}$
is compared to the experimental pdf (dots) in linear 
scales.}
\label{fig8}
\end{figure}

It is important to emphasize that the remarkable fittings shown in Figs. 5-8, between the numerical and experimental pdfs for 
the set $\{ s_{11}, s_{ij} \}$, are obtained from the mapping, determined by the single parameter $N_A - N_B$, provided by the 
fluctuations given by (\ref{eq2.25}). This constitutes strong evidence for the existence of an underlying log-Poisson 
cascade process. We note, furthermore, that the agreement between modelled and experimental pdfs 
would be not so good if the experimental pdfs of $s_{22}$ or $s_{33}$ were chosen in place of the 
one for $s_{11}$. The present method, thus, has the heuristic potential to address issues of isotropy 
in boundary layer flows.

A further application of our results is the definition of an {\it{effective}} Reynolds number $\bar R_\lambda$ 
for the atmospheric surface turbulent flow, taking the more controlled Reynolds number of the DNS as a standard. 
We write, according to (\ref{hyperfb}),
\be
\bar R_\lambda = 240 \times \left ( \frac{11.5}{6.6} \right )^3 \simeq 1.2 \times 10^3 \ . \ \label{eq4.3}
\ee
It was noted, in Ref. \cite{guli1_etal}, that the rough estimate $R_\lambda = 3.4 \times 10^3$ displaces the
point $(R_\lambda, H_4) = (3.4 \times 10^3, 11.5)$ out of the empirical curve well modelled 
$H_4 \sim R_\lambda^{\alpha_4}$. However, we find that if the alternative value (\ref{eq4.3}) is used instead of 
$R_\lambda = 3.4 \times 10^3$, then the point $(R_\lambda, H_4)$ gets closer to the usual curve of flatness.

\section{Conclusions}

We have used the log-Poisson model of the turbulent cascade to get the pdfs of velocity 
gradient fluctuations of a high Reynolds turbulent atmospheric flow. The excellent fittings are
achieved by means of a Monte-Carlo integration procedure and the use of standard pdfs obtained 
in a lower Reynolds number DNS. Our results indicate that non-gaussianity and anomalous scaling of
scale dependent observables can be seen as different manifestations of intermittency that can be 
approached within a unified framework. Actually, this point of view has been formerly pursued 
along the multifractal description of intermittency \cite{benzi1}, with modest success in the 
quality of pdf fittings, nevertheless the fact that they are dependent on a large number of free 
parameters. 

As a natural application of our methodology, we have found a way to (i) select isotropy sectors 
of the velocity gradient tensor in boundary layer flows and (ii) unambiguously define effective
Taylor-based Reynolds numbers in the presence of anisotropy. These results can be of considerable
interest in the study of anisotropy effects in turbulent boundary layers. It is also likely that
the same ideas can be extended to the case of free shear turbulence.

An interesting question is how low can be the DNS Reynolds number, while still leading to good
velocity gradient pdf fittings for higher Reynolds number flows, along the lines discussed
in Sec. IV. An investigation of this matter could throw some light on the problem 
of extended self-similarity \cite{benzi2}. Also, we wonder if correlation effects in the velocity 
gradient time series could be modelled in similar ways. A promising direction here would be to 
link the Fokker-Planck approach to turbulent time series \cite{peinke} with the log-Poisson 
cascade model.

It is clear that the multiplicative cascade picture is worth as a phenomenological construction if 
a consistent meaning can be given to concepts like the inertial range, local cascade, and the 
universality of velocity structure exponents. However, recent work  \cite{kholmy2} on the scaling 
behavior of velocity structure functions suggests that inertial and dissipative range fluctuations 
could be coupled in a bidirectional way. It has been found in \cite{kholmy2} that the scaling 
exponents measured in the inertial range are changed if strong dissipative events are discarded in 
the averaging procedure, indicating a ``flow of influence" from the small to the large scales. 

In order to address further related studies, we note that a possible solution 
to these puzzling observations, saving the essence of the multiplicative cascade phenomenology, would 
rely on the usual definition of the energy dissipation rate $\epsilon_m$ as the local dissipation rate 
averaged over volumes with linear sizes of the order of $\ell_m = L/a^m$. Since the energy dissipation 
rate is long-range correlated, it is likely that events which have strong local dissipation rates
turn to be correlated to strong events in the above (inertial range averaged) sense. 

\acknowledgments

We thank the US-Israel Binational Foundation and the Israel Science Foundation (M.K. and A.T.) and CAPES, 
CNPq and FAPERJ (L.M. and R.M.P.) for support. We are mostly indebted to P.K. Yeung for his kind attention 
in providing us with DNS data and to F.S. Amaral for interesting discussions.


\begin{references}
\bibitem{batch-town} G.K. Batchelor and A.A. Townsend, Proc. R. Soc. Lond. A {\bf{199}},
238 (1949).
\bibitem{k41a} A.N. Kolmogorov, Dokl. Akad. Nauk SSSR {\bf{30}}, 9 (1941); english translation:
Proc. Roy. Soc. London A {\bf{434}}, 9 (1991).
\bibitem{k41b} A.N. Kolmogorov, Dokl. Akad. Nauk SSSR {\bf{32}}, 16 (1941); english translation:
Proc. Roy. Soc. London A {\bf{434}}, 15 (1991).
\bibitem{frisch2} U. Frisch, {\it{Turbulence: The Legacy of A.N. Kolmogorov}}, Cambridge University Press, 
Cambridge (1995).
\bibitem{frisch1} U. Frisch, Proc. R. Soc. Lond. A {\bf{434}},
99 (1991).
\bibitem{ansel_etal} F. Anselmet, Y. Gagne, E.J. Hopfinger e R.A. Antonia, J. Fluid Mech. {\bf{140}}, 
63 (1984).
\bibitem{k62} A.N. Kolmogorov, J. Fluid Mech. {\bf{13}}, 82 (1962).
\bibitem{ob62} A.M. Obukhov, J. Fluid Mech. {\bf{13}}, 77 (1962).
\bibitem{tsinober} A. Tsinober, {\it{An Informal Introduction to Turbulence}}, Kluwer Academic Press, Netherlands (2001).
\bibitem{bec} J. Bec and K. Khanin, Phys. Rep. {\bf{447}}, 1 (2007).
\bibitem{dubrulle} B. Dubrulle, Phys. Rev. Lett. {\bf{73}}, 969 (1994).
\bibitem{she} Z-S. She and E.C. Waymire, Phys. Rev. Lett. {\bf{74}}, 262 (1995).
\bibitem{she-leveque} Z-S. She e E. Leveque, Phys. Rev. Lett. {\bf{72}}, 336 (1994).
\bibitem{guli1_etal} G.Gulitski, M. Kholmyansky, W. Kinzelbach, B. L\"uthi, A. Tsinober and S. Yorish,
J.Fluid Mech. {\bf 589}, 57 (2007).
\bibitem{guli2_etal} G.Gulitski, M. Kholmyansky, W. Kinzelbach, B. L\"uthi, A. Tsinober and S. Yorish,
J.Fluid Mech. {\bf 589}, 83 (2007).
\bibitem{guli3_etal} G.Gulitski, M. Kholmyansky, W. Kinzelbach, B. L\"uthi, A. Tsinober and S. Yorish,
J.Fluid Mech. {\bf 589}, 103 (2007).
\bibitem{donzis} D. Donzis, P.K. Yeung and K. Sreenivasan, Phys. Fluids {\bf{20}}, 045108 (2008).
\bibitem{kholmy} M. Kholmyansky, L. Moriconi and A. Tsinober, 
Phys. Rev. E {\bf{76}}, 026307 (2007).
\bibitem{comment} The use of log-normal statistics in Ref. \cite{kholmy} is now understood to be
just a good approximation to the improved picture provided by the log-Poisson model.
\bibitem{wyn-tenn} J.C. Wyngaard and H. Tennekes, Phys. Fluids {\bf 13}, 1962 (1970).
\bibitem{gagne} Y. Gagne, M. Marchand and B. Castaing, J. Phys. II France {\bf{4}}, 1 (1994).
\bibitem{naert} A. Naert, B. Castaing, B. Chabaud, B. Htbral and J. Peinke, Physica D {\bf{113}}, 73 (1998).
\bibitem{ishi_etal} T. Ishihara, Y. Kaneda, M. Yokokawa, K. Itakura and A. Uno, J. Fluid Mech. {\bf{592}}, 335 (2007).
\bibitem{knuth} D. Knuth, {\it{The Art of Computer Programming: V. 2 - Seminumerical Algorithms}}, 
Addison Wesley, Reading, Massachusets (1981).
\bibitem{mc}  M.E.J. Newman and G.T. Barkema, {\it{Monte Carlo Methods in Statistical Physics}}, 
Oxford University Press, Oxford (2001).
\bibitem{benzi1} R. Benzi, L. Biferale, G. Paladin, A. Vulpiani and M. Vergassola, Phys. Rev. Lett. {\bf{67}},
2299 (1991).
\bibitem{benzi2} R. Benzi, S. Ciliberto, R. Tripiccione, C. Baudet, F. Massaioli e S. Succi, 
Phys. Rev. E {\bf{48}}, R29 (1995).
\bibitem{peinke} J. Peinke, A. Nawoth, S.T. L\"uck, M. Siefert and R. Friedrich, {\it{Stochastic Analysis and
New Insights into Turbulence}}, Advances in Turbulence XI, Springer (2007).
\bibitem{kholmy2} M. Kholmyansky and A. Tsinober, Phys. Lett. A {\bf{37}}, 2364 (2009).
\end{references}
\end{document}